\documentclass{article}

\usepackage{arxiv}

\usepackage[utf8]{inputenc} 
\usepackage[T1]{fontenc}    
\usepackage{hyperref}       
\usepackage{url}            
\usepackage{booktabs}       
\usepackage{amsfonts}       
\usepackage{nicefrac}       
\usepackage{microtype}      
\usepackage{lipsum}		
\usepackage{graphicx}
\usepackage[numbers]{natbib}
\usepackage{doi}
\usepackage{amsmath,amssymb}

\title{Different Cybercrimes and their Solution for Common People}

\author{ \href{https://orcid.org/0009-0007-8038-1278}{\includegraphics[scale=0.06]{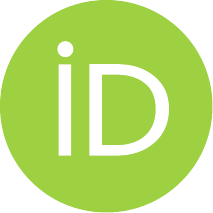}\hspace{1mm}Sagar Tamang}\thanks{Correspondance can be addressed to \textit{cs22bcagn033@kazirangauniversity.in}} \\
	Department of Information Technology\\
	The Assam Kaziranga University\\
	Jorhat, India \\
	\texttt{cs22bcagn033@kazirangauniversity.in} \\
	\And
        {G. Sai chandana}\\
	Department of Electrical Engineering\\
	MITS, Madanapalle\\
	Madanapalle, India \\
	\texttt{sai772704@gmail.com} \\
	\And
	\href{https://orcid.org/0000-0003-2060-3282}{\includegraphics[scale=0.06]{orcid.pdf}\hspace{1mm}Binoy Krishna Roy} \\
	Department of Electrical Engineering\\
	National Institute of Technology Silchar\\
	Silchar, India \\
	\texttt{bkr@ee.nits.ac.in} \\
}

\renewcommand{\shorttitle}{Different Cybercrimes \& their Solution for Common People}

\hypersetup{
pdftitle={Different Cybercrimes and their Solution for Common People},
pdfsubject={cs.CR},
pdfauthor={S. Tamang, G. S. Chandana, B. K. Roy},
pdfkeywords={cyberspace, cybercrime, solutions, common man},
}

\begin{document}
\maketitle

\begin{abstract}
In today's digital age, cyberspace has become integral to daily life, however it has also led to an increase in cybercriminal activities. This paper explores cybercrime trends and highlights the need for cybercrime awareness (cyberawareness) to mitigate vulnerabilities. The study also examines Indian statistics on cybercrime. We review the existing literature on cybercrime and cybersecurity, focusing on various types of cybercrimes and their impacts. We present a list of 31 technical as well as 
Common man solutions, considering that they are not technologically updated. Expanding the list of solutions and validating their effectiveness in cyber threats can be the future scope of the research.
\end{abstract}


\section{Introduction}
\label{sec:introduction}
In today's digital civilization, cyberspace (also known as the internet world) is used extensively. However, with an ever-increasing digital footprint, cybercriminal activities have also increased \cite{Humayun2020Cyber,cybercrimeinindiaempericalstudy}. Therefore, the topic of cybersecurity has become a topic of great interest for researchers and practitioners \cite{Humayun2020Cyber}. Cybersecurity is when a crime is committed where a computer is used as a tool or when a computer is subject to a crime \cite{cybercrimeinindiaempericalstudy}. 

The vast majority of political, financial, and military organizations, among many others, process a great deal of sensitive \cite{emergingcybersecurity,idtheft2023}. As such, organizations and Governments around the globe are racing towards research in the field of cyber security to protect themselves against any emerging or future threats \cite{mappingcybersecurityresearch}. According to recent studies \cite{idtheft2023} in the United States, healthcare has become the top industry to be compromised, among many others. 

The rest of the paper is structured in the following way: Section \ref{sec:literature} introduces other relevant studies in the field; Section \ref{sec:solution} provides our solutions; Section \ref{sec:discussion} is where the discussions are conducted; Section \ref{sec:conclusions} gives a conclusion of the research with future outlook.

\section{Literature Review}
\label{sec:literature}
\subsection{Cybercrime}
M Humayun et al. conducted a systematic mapping study of 66 primary studies to identify common cybersecurity vulnerabilities. Their research concluded that most selected studies target only a few common security vulnerabilities such as phishing, denial of service, and malware \cite{Humayun2020Cyber}.

J Srinivas et al. discussed various aspects of cybersecurity, including cyber attacks, security requirements, and the architecture of cybersecurity incident management frameworks (CIMF). They emphasized the importance of standardization in cybersecurity and outlined challenges in implementing these standards. Their paper also reviewed government policies and provided recommendations for improving cybersecurity measures and defence strategies \cite{Srinivas2019Government}.

V Sharma et al. examined the surge in cyber-crimes amid rapid technological growth. They categorized cybercrimes, such as hacking and phishing, and highlighted the lack of significant correlation between demographics and awareness. Based on a survey of 134 residents in Delhi and NCR, their study underscored the need for enhanced public awareness and security measures to combat cybercrime in the digital age \cite{growthofcybercrimesinsociety}.

In his paper, "Cyber Crime in India: An Empirical Study," S Ahmad provided an extensive overview of cybercrime and its legal framework in India. The study highlighted the increasing prevalence of cybercrime, a phenomenon where computers are used as tools or targets for illegal activities. The paper discussed the evolution of cybercrime laws in India, notably the Information Technology Act of 2000, which addressed e-commerce, electronic governance, and penalties for computer-related crimes \cite{cybercrimeinindiaempericalstudy}.   

P Shah and A Agarwal investigated the cybersecurity behaviour of smartphone users in India. Their study identified 28 key cybersecurity practices through a literature review and surveyed 300 users. The research found that while users adopt basic security measures like screen locks, they often neglect advanced controls such as encryption. Significant differences in security practices were observed based on factors like gender, age, and mobile OS. The study highlighted a gap in advanced cybersecurity practices and emphasised the need for targeted intervention programs for Indian smartphone users \cite{cybersecuritybehaviourofsmartphoneusers}.

Bahuguna et al. evaluated the cybersecurity maturity of Indian organizations, highlighting significant gaps despite national efforts like the Cyber Essentials program and India's National Cybersecurity Policy 2013. Their research underscored challenges such as limited budgets, insufficient skilled personnel, and inadequate management support, aligning with broader findings that emphasize the need for robust policies, enhanced technical controls, and increased capacity building to effectively address cybersecurity threats \cite{assessingcybersecuritymaturity}.

In reviewing the landscape of Indian cybersecurity research from 1999 to 2020, Elango et al. conducted a scientometric analysis based on the Scopus database. Their study revealed a significant increase in publication output, especially in recent years, with a notable shift from data-centric topics to emerging areas such as machine learning, cloud computing, and artificial intelligence. They found that conference papers are the predominant publication type and identified key authors, institutions, and international collaborations. Their analysis positioned India as a major player in the field, ranking fourth globally in research output, underscoring the country’s evolving focus in cybersecurity research \cite{mappingcybersecurityresearch}.

\subsection{Types of Cybercrimes}
We have handpicked a list of 31 common cybercrimes for our study \cite{cybercrimeinindiaempericalstudy,NCRB2018Crime2,NCRB2019Crime2,NCRB2020Crime2,NCRB2021Crime2,NCRB2022Crime2}, and the list is as follows:

\begin{enumerate}
\item \textbf{Phishing}: Deceptive attempts to obtain sensitive information by disguising it as a trustworthy entity.

\item \textbf{Identity Theft}: Unauthorized use of someone else's personal information for fraud.

\item \textbf{Hacking}: Unauthorized access to or manipulation of computer systems or networks.

\item \textbf{Ransomware}: Malware that encrypts data and demands payment for its release.

\item \textbf{Cyberstalking}: Repeated use of electronic communications to harass or intimidate an individual.

\item \textbf{Online Fraud}: Deceptive online practices aimed at financial gain, such as scams and false advertising.

\item \textbf{Cyberbullying}: Use of digital platforms to harass, threaten, or embarrass individuals, often repeatedly.

\item \textbf{DDoS Attacks}: Distributed Denial-of-Service attacks overwhelm a network with traffic, causing it to crash.

\item \textbf{Malware}: Malicious software designed to disrupt, damage, or gain unauthorized access to systems.

\item \textbf{Spyware}: Software that secretly monitors and collects user information without consent.

\item \textbf{Adware}: Unwanted software designed to display advertisements, often leading to further malware.

\item \textbf{Social Engineering}: Manipulating individuals into divulging confidential information through deception.

\item \textbf{Online Scams}: Fraudulent schemes conducted over the internet, often targeting vulnerable individuals.

\item \textbf{Botnets}: Compromised computer networks controlled by attackers to conduct large-scale attacks.

\item \textbf{Cryptojacking}: Unauthorized use of someone else's computing power to mine cryptocurrency.

\item \textbf{Credential Stuffing}: Automated injection of breached username/password pairs to gain unauthorized access.

\item \textbf{Eavesdropping Attacks}: Intercepting communications between parties to steal data or secrets.

\item \textbf{Online Extortion}: Coercing individuals or organizations into making payments under threats.

\item \textbf{Cyber Espionage}: Unauthorized access and theft of information from government or corporate entities for strategic advantage.

\item \textbf{Fake Online Marketplaces}: Fraudulent e-commerce sites created to steal payment information.

\item \textbf{Carding}: The trafficking and unauthorized use of credit card information.

\item \textbf{Romance Scams}: Fraudsters posing as romantic interests to defraud victims of money.

\item \textbf{Business Email Compromise}: Fraudulent attempts to trick businesses into transferring money or data via email.

\item \textbf{Internet Piracy}: Unauthorized distribution and use of copyrighted digital content.

\item \textbf{Child Exploitation}: Using the internet to abuse or exploit children, including the distribution of illegal content.

\item \textbf{Revenge Porn}: Sharing private, explicit images or videos of someone without their consent.

\item \textbf{Fake News and Misinformation}: The intentional spread of false information online to mislead or manipulate public opinion.

\item \textbf{Insider Threats}: Security risks that originate within an organization, often by disgruntled or compromised employees.

\item \textbf{IoT Device Hacking}: Compromising Internet of Things (IoT) devices to steal data or launch attacks.

\item \textbf{Data Breaches}: Unauthorized access and disclosure of sensitive, protected, or confidential data.

\item \textbf{Zero-Day Exploits}: Attacks that exploit previously unknown vulnerabilities in software or hardware.
\end{enumerate}

Table \ref{table:cybercrimesexample} presents notable examples of all 31 common cybercrimes. Each of these examples is mostly taken from the Indian context.

\begin{figure}
    \centering
    \includegraphics[width=1\linewidth]{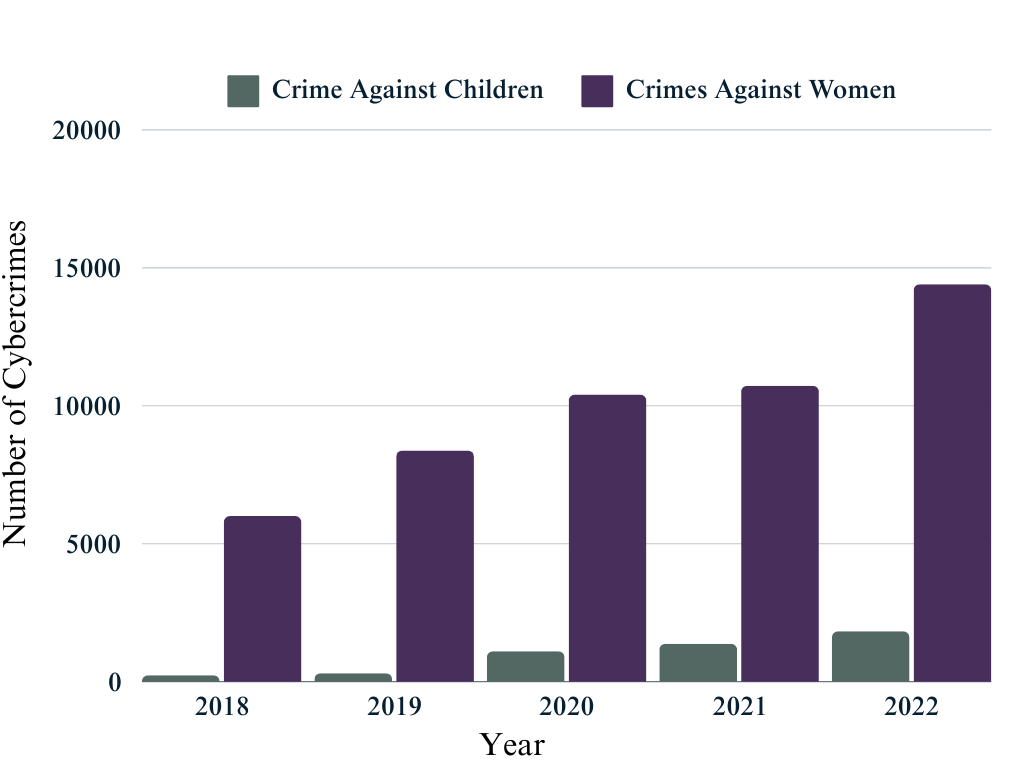}
    \caption{Cybercrimes Against Women and Children in India (in Numbers) \cite{NCRB2018Crime2,NCRB2019Crime2,NCRB2020Crime2,NCRB2021Crime2,NCRB2022Crime2}}
    \label{fig:crimes-women-children}
\end{figure}

\begin{figure}
    \centering
    \includegraphics[width=1\linewidth]{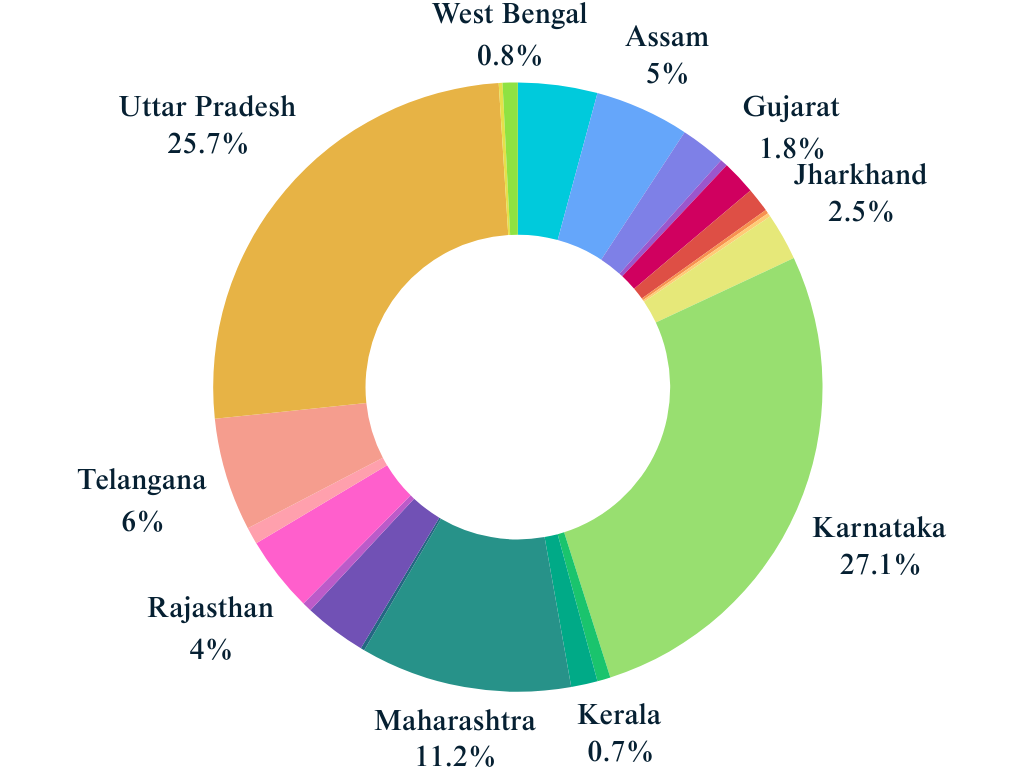}
    \caption{Visual Distribution of Cybercrime Cases Across Indian States (\%) \cite{NCRB2018Crime2,NCRB2019Crime2,NCRB2020Crime2,NCRB2021Crime2,NCRB2022Crime2}}
    \label{fig:crimes-in-india-state}
\end{figure}

\subsection{General Trends of Cybercrime}
The National Crime Records Bureau (NCRB), under the Ministry of Home Affairs, the Government of India, is a very good source for statistics on crimes in India, including cybercrimes. Much of the volume 2 of the NCRB has many sections dedicated entirely to cybercrimes. One disadvantage of NCRB's data is that there is no latest data after 2022.

As can be observed in Figure \ref{fig:crimes-women-children}, there has been a steady increase in cybercrimes against women and children. Figure \ref{fig:crimes-in-india-state} and Table \ref{table:state-ut-cybercrimes} visually and textually present the percentage per state of cybercrimes in India for the year 2018 \cite{NCRB2018Crime2,NCRB2019Crime2,NCRB2020Crime2,NCRB2021Crime2,NCRB2022Crime2}. The NCRB's data stopped publishing the percentage per state data from 2020 onwards, hence the 2018 data. It can be observed that most of the cybercrimes are centred around the states of Karnataka and Uttar Pradesh, with 27.1\% and 25.7\%, respectively, making up more than 50\% of the country (52.8\% to be precise). The assumption we can make for this is due to the presence of tier-1 cities in these states: Bangalore and Noida \cite{consumertiercity}, where a lot of the cyber/internet activity is present. Bangalore was the first city in India to have a Cybercrime Police Station \cite{KarnatakaCyberCrime}.

\begin{table}[h!]
\centering
\caption{Cybercrimes in India (\% per state) \cite{NCRB2018Crime2,NCRB2019Crime2,NCRB2020Crime2,NCRB2021Crime2,NCRB2022Crime2}}
\begin{tabular}{|l|r|}
\hline
\textbf{State/UT}        & \textbf{Percentage} \\ \hline
Andhra Pradesh           & 4.2\%               \\ \hline
Arunachal Pradesh        & 0.0\%               \\ \hline
Assam                    & 5.0\%               \\ \hline
Bihar                    & 2.4\%               \\ \hline
Chhattisgarh             & 0.4\%               \\ \hline
Goa                      & 0.0\%               \\ \hline
Gujarat                  & 1.8\%               \\ \hline
Haryana                  & 1.3\%               \\ \hline
Himachal Pradesh         & 0.2\%               \\ \hline
Jammu \& Kashmir         & 0.2\%               \\ \hline
Jharkhand                & 2.5\%               \\ \hline
Karnataka                & 27.0\%              \\ \hline
Kerala                   & 0.7\%               \\ \hline
Madhya Pradesh           & 1.4\%               \\ \hline
Maharashtra              & 11.2\%              \\ \hline
Manipur                  & 0.0\%               \\ \hline
Meghalaya                & 0.2\%               \\ \hline
Mizoram                  & 0.0\%               \\ \hline
Nagaland                 & 0.0\%               \\ \hline
Odisha                   & 3.3\%               \\ \hline
Punjab                   & 0.5\%               \\ \hline
Rajasthan                & 4.0\%               \\ \hline
Sikkim                   & 0.0\%               \\ \hline
Tamil Nadu               & 0.9\%               \\ \hline
Telangana                & 6.0\%               \\ \hline
Tripura                  & 0.0\%               \\ \hline
Uttar Pradesh            & 25.6\%              \\ \hline
Uttarakhand              & 0.2\%               \\ \hline
West Bengal              & 0.8\%               \\ \hline
\end{tabular}
\label{table:state-ut-cybercrimes}
\end{table}

\label{sec:solution}
\begin{table*}[h!]
    \centering
    \scriptsize
    \caption{Different cybercrimes and notable incidents associated with them.}
    \begin{tabular}{|c|l|p{11cm}|}
        \hline
        \textbf{Index} & \textbf{Name} & \textbf{Example} \\
        \hline
        1 & Phishing & Union Bank of India Heist: Hackers initiated a fraudulent transfer of \$171 million from Union Bank of India to various offshore accounts \cite{livemint2016unionbank}. \\
        \hline
        2 & Identity Theft & Aadhaar Data Breach (India): Personal information of over 1.1 billion Indian citizens was exposed due to a security flaw in the Aadhaar system \cite{uw2024aadhaarcybersecurity}. \\
        \hline
        3 & Hacking & Sony Pictures Hack: In 2014, North Korean hackers infiltrated Sony's network, leaking sensitive data and unreleased films \cite{northkoreasonyhack}. \\
        \hline
        4 & Ransomware & WannaCry Attack: A global ransomware attack in 2017 affected over 200,000 computers across 150 countries, including India's Andhra Pradesh police department \cite{cloudflare_wannacry}. \\
        \hline
        5 & Cyberstalking & Ritu Kohli Case (2003): India's first reported cyberstalking case, where Ritu Kohli's details were posted online without her consent by her husband's friend \cite{deo2013cyberstalking}. \\
        \hline
        6 & Online Fraud & Nigerian Prince Scam: A common online fraud where victims are promised a share of a large sum of money in exchange for personal information and financial assistance \cite{nigerianprince}. \\
        \hline
        7 & Cyberbullying & Ooshmal Ullas (2016): A 23-year-old MBBS student from Kerala, Ooshmal Ullas, tragically committed suicide after facing severe cyberbullying related to a Facebook post \cite{Kaur2023}. \\
        \hline
        8 & DDoS Attacks & Government Websites Targeted (2023): In May 2023, hacktivist groups claimed responsibility for 480 DDoS attacks on various Indian government websites \cite{Radware2023}. \\
        \hline
        9 & Malware & AIIMS Delhi Ransomware Attack (2023): A ransomware attack at AIIMS Delhi disrupted health services and may have compromised patient data \cite{DSCI2023}. \\
        \hline
        10 & Spyware & Pegasus Spyware: On July 10, 2024, Apple issued two fresh pegasus spyware alerts in India, warning them that they had been targeted by mercenaries \cite{hindu2024pegasus}. \\
        \hline
        11 & Adware & Mobile Adware: Adware can get onto people's mobile phones through seemingly harmless apps in categories like entertainment and gaming \cite{KasperskyAdware}. \\
        \hline
        12 & Social Engineering & A retired IAS officer was scammed out of INR 1.89 crore (approximately \$250,000) by fraudsters posing as forex trading experts \cite{Bureau2023}. \\
        \hline
        13 & Online Scams & With the rise of UPI in India, scammers exploit the system by tricking victims into sharing PINs or clicking on malicious links that lead to unauthorized transactions \cite{IndiaToday2024}. \\
        \hline
        14 & Botnets & Mirai Botnet: In 2016, this botnet took down major websites like Twitter and Netflix, affecting users globally, including India \cite{ForumIAS2023}. \\
        \hline
        15 & Cryptojacking & Cryptojacking of Government Websites (India): Several Indian government websites were hijacked to mine cryptocurrency in 2018 \cite{CoinDesk2018}. \\
        \hline
        16 & Credential Stuffing & Uber Credential Stuffing Attack (2018): In a high-profile incident, Uber faced a credential stuffing attack that compromised the personal information of approximately 82,000 drivers and 2.7 million customers \cite{NDTVProfit2024}. \\
        \hline
        17 & Eavesdropping Attacks & Pegasus Spyware Allegations (2021): Pegasus spyware allegedly targeted opposition leaders, journalists, and activists in India, raising major privacy concerns \cite{AlJazeera2023} \\
        \hline
        18 & Online Extortion & Bengaluru Ransomware Attack (2023): BSR Infratech in Bengaluru was hit by ransomware demanding \$80,000 (approximately INR 66.7 lakh) for decryption. Hackers threatened to sell the data on the dark web if not paid \cite{DeccanHerald2023}. \\
        \hline
        19 & Cyber Espionage & Kudankulam Nuclear Plant Malware: The North Korean Lazarus Group's Dtrack malware targeted Kudankulam, aiming to collect confidential data and raise security concerns \cite{Kratikal2023}. \\
        \hline
        20 & Fake Online Marketplaces & Counterfeit Products on Major E-commerce Platforms (2022): A survey revealed that many consumers reported purchasing counterfeit products from well-known e-commerce platforms \cite{EUHelpdesk2022}. \\
        \hline
        21 & Carding & Gwalior Teen's Carding Operation (2020): A 17-year-old boy from Gwalior was arrested for using stolen international credit card data to make purchases worth INR 1.5 crore (approximately \$200,000) \cite{TimesOfIndia2020}. \\
        \hline
        22 & Romance Scams & Tinder Swindler: A scammer posed as a wealthy individual on Tinder, defrauding victims in India and worldwide by convincing them to transfer money \cite{IndiaToday2024-2}. \\
        \hline
        23 & Business Email Compromise & Wipro Email Compromise: In 2019, Wipro's email system was hacked, leading to fraudulent invoices being sent to clients \cite{ComputerWeekly2024}. \\
        \hline
        24 & Internet Piracy & Tamilrockers: A notorious Indian website known for leaking pirated copies of movies and TV shows, affecting the entertainment industry \cite{EconomicTimes2024}. \\
        \hline
        25 & Child Exploitation & Operation Megh Chakra (India): A nationwide crackdown on child pornography resulted in the arrest of individuals sharing exploitative content online \cite{TheHindu2022}. \\
        \hline
        26 & Revenge Porn & Case of Nagpur Woman: A woman in Nagpur was a victim of revenge porn when her ex-boyfriend posted private photos to 11 people including her husband and some co-workers \cite{FreePressJournal2023}. \\
        \hline
        27 & Fake News \& Misinformation & COVID-19 Misinformation: During the pandemic, fake news about cures and treatments spread widely on social media, causing public confusion and panic \cite{TimesOfIndia2021}. \\
        \hline
        28 & Insider Threats & Tesla Insider Sabotage: In 2018, a disgruntled employee sabotaged Tesla’s manufacturing system, leaking sensitive data \cite{SCMagazine2023}. \\
        \hline
        29 & IoT Device Hacking & Mirai IoT Botnet: Infected IoT devices worldwide, including India, to launch large-scale DDoS attacks \cite{Cloudflare2024}. \\
        \hline
        30 & Data Breaches & Aadhaar Data Leak: In 2018, a breach exposed the personal data of over 1.1 billion Indian citizens registered in the Aadhaar database \cite{DrishtiIAS2024}. \\
        \hline
        31 & Zero-Day Exploits & Stuxnet Worm: A sophisticated zero-day exploit targeting Iran’s nuclear facilities, with ripple effects on industrial systems worldwide \cite{Holloway2015}. \\
        \hline
        32 & Digital Arrest &  A call is established by cyber criminals. They create an environment as if victims are involved in illegal activities. Other criminals appear on the scene as members of some investigating agencies and offer help. They force the victims to keep their video on and to follow their instructions. Victims follow the instructions and become victims of digital arrest \cite{the_Hindu_16th_August_2024}.    \\
        \hline
    \end{tabular}
    \label{table:cybercrimesexample}
\end{table*}

\subsection{Significance of Cybercrime Awareness}
According to Curtis and Colwell, the risk of cybercrimes decreases proportionately to the more aware they are. The risk in cyberspace can be reduced by educating young people about cybercrimes \cite{curtis2000cyber,growthofcybercrimesinsociety}.

\section{Solutions of Common Cybercrimes}
In light of these studies, we propose 31 technical and non-technical solutions for 31 common cybercrimes. The solutions are available in Table \ref{table:cybercrime-solutions}.

\begin{table*}[h!]
\centering
\scriptsize
\caption{Non-Technical and Technical Solutions for 31 Common Cybercrimes}
\begin{tabular}{|c|l|p{5cm}|p{5cm}|}
\hline
\textbf{Index} & \textbf{Name} & \textbf{Non-Technical Solution (Simple)} & \textbf{Technical Solution} \\ \hline
1 & Phishing & Be cautious of emails or messages asking for personal information. & Use email filtering, anti-phishing tools, and multi-factor authentication. \\ \hline
2 & Identity Theft & Regularly check bank statements and credit reports for unusual activity. & Implement identity theft protection services and secure personal information. \\ \hline
3 & Hacking & Use strong, unique passwords for different accounts. & Enable firewalls, use intrusion detection systems, and update software regularly. \\ \hline
4 & Ransomware & Back up important files on an external device or cloud storage. & Use anti-ransomware software, patch vulnerabilities, and employ network segmentation. \\ \hline
5 & Cyberstalking & Block and report online harassment immediately. & Use privacy settings on social media, employ VPNs, and track IP addresses. \\ \hline
6 & Online Fraud & Verify the authenticity of websites before making purchases. & Use secure payment gateways and fraud detection software. \\ \hline
7 & Cyberbullying & Talk to someone you trust if you or someone you know is being bullied online. & Use content filtering and monitoring tools and report abusive behaviour on platforms. \\ \hline
8 & DDoS Attacks & If a website is down, wait and try again later. & Implement DDoS protection services and rate-limiting on servers. \\ \hline
9 & Malware & Avoid downloading software or files from unknown or untrusted sources. & Install and update antivirus software, use application whitelisting, and enable firewalls. \\ \hline
10 & Spyware & Be cautious with free software and avoid clicking on suspicious ads. & Use anti-spyware tools, regularly scan your system, and employ sandboxing techniques. \\ \hline
11 & Adware & Avoid downloading apps from unofficial or unknown sources. & Use ad blockers and antivirus software to detect and remove adware. \\ \hline
12 & Social Engineering & Don’t share personal information with strangers who contact you unexpectedly. & Implement multi-factor authentication and conduct security awareness training. \\ \hline
13 & Online Scams & Be sceptical of deals that seem too good to be true, especially online. & Use scam detection tools, secure browsing solutions, and educate users about common scams. \\ \hline
14 & Botnets & Keep your software and devices updated regularly. & Implement network monitoring and endpoint security solutions to detect botnet activity. \\ \hline
15 & Cryptojacking & If your computer is running slow, check for unusual activity. & Use anti-crypto jacking browser extensions and endpoint protection software. \\ \hline
16 & Credential Stuffing & Use different passwords for different accounts. & Implement account lockout mechanisms and multi-factor authentication. \\ \hline
17 & Eavesdropping Attacks & Avoid using public Wi-Fi for sensitive activities. & Use VPNs, encryption protocols, and secure communication channels. \\ \hline
18 & Online Extortion & Do not pay; report the incident to the authorities immediately. & Use data encryption, access controls, and regular backups to protect sensitive data. \\ \hline
19 & Cyber Espionage & Be cautious about opening suspicious emails or files from unknown sources. & Use advanced threat detection tools, encryption, and secure communications. \\ \hline
20 & Fake Online Marketplaces & Only shop on well-known and trusted websites. & Use website verification tools and secure payment methods to prevent fraud. \\ \hline
21 & Carding & Monitor your bank statements for unexpected charges, even small ones. & Use fraud detection and prevention services, as well as secure payment systems. \\ \hline
22 & Romance Scams & Be careful if someone you meet online quickly asks for money or personal information. & Use background checks and online safety tools, and educate users on romance scam tactics. \\ \hline
23 & Business Email Compromise & Verify unusual or unexpected requests for money transfers with a phone call. & Implement email authentication protocols, encryption, and employee training programs. \\ \hline
24 & Internet Piracy & Use legal streaming services or purchase digital content through authorized vendors. & Implement digital rights management (DRM) solutions and anti-piracy software. \\ \hline
25 & Child Exploitation & Monitor your child's online activities and talk to them about online safety. & Use parental control software and report suspicious activities to authorities. \\ \hline
26 & Revenge Porn & Report the content to the platform and seek legal assistance immediately. & Use automated content monitoring, digital forensics, and privacy protection tools. \\ \hline
27 & Fake News and Misinformation & Verify news and information from multiple reliable sources before sharing. & Use fact-checking algorithms, content moderation tools, and promote media literacy. \\ \hline
28 & Insider Threats & If you notice suspicious behavior at work, report it to a supervisor or HR. & Implement access controls, employee monitoring systems, and conduct security audits. \\ \hline
29 & IoT Device Hacking & Change the default passwords on your smart home devices and update them regularly. & Use network segmentation, IoT security solutions, and device firmware updates. \\ \hline
30 & Data Breaches & Use strong passwords and be cautious about sharing personal information online. & Implement encryption, data loss prevention (DLP) tools, and regular security audits. \\ \hline
31 & Zero-Day Exploits & Regularly update your software and install patches as soon as they are available. & Use advanced threat protection, zero-day exploit detection, and patch management systems. \\ \hline
32 & Digital Arrest & Awareness on the role of any government investigating agencies & Discuss the matter with family members before the transfer of huge funds and immediately inform the cybercrime authority after such a crime. \\ \hline
\end{tabular}
\label{table:cybercrime-solutions}
\end{table*}

\section{Discussion}
\label{sec:discussion}
Our technical and non-technical solutions presented in Table \ref{table:cybercrime-solutions} encompass many common cybercrimes (31 to be precise) but it does not capture all the cybercrimes, this can be where future studies could be conducted. In Table \ref{table:state-ut-cybercrimes}, the vast majority of cybercrimes are centred around the states of Karnataka, Uttar Pradesh, and Maharashtra. The reason for this could further be investigated as a future study. In our research, as seen in Figure \ref{fig:crimes-women-children}, we have found data for specific numbers of crimes against women and children; this can be further expanded for better insights and clarifications. The solutions we offer can be researched to see if these bring about a positive change in society, showcasing a possible future research scope. 

\section{Conclusions}
\label{sec:conclusions}
In this research, we examined cybersecurity trends and the importance of cyber awareness in mitigating vulnerabilities. We presented 31 Technical as well as Non-Technical Solutions for 31 common cybercrimes. Spreading awareness of these 31 solutions can help the common man be more secure from growing cybercrimes. Future research could focus on expanding these solutions or validating the effectiveness of the proposed measures.

\section{Acknowledgments}
\label{sec:acknowledgments}
The authors would like to thank the IEEE Subsection,  National Institute of Technology Silchar, for allowing them to do this work under a summer internship program. The first author thanks Assam Kaziranga University for sponsoring him for this work. The second author expresses her thanks to Madanapalle Institute of Technology and Sciences for permitting her to join this internship program.

\end{document}